\documentstyle[12pt,axodraw]{article}
\begin{document}
\baselineskip 18pt
\def\today{\ifcase\month\or
 January\or February\or March\or April\or May\or June\or
 July\or August\or September\or October\or November\or December\fi
 \space\number\day, \number\year}
\def\thebibliography#1{\section*{References\markboth
 {References}{References}}\list
 {[\arabic{enumi}]}{\settowidth\labelwidth{[#1]}
 \leftmargin\labelwidth
 \advance\leftmargin\labelsep
 \usecounter{enumi}}
 \def\newblock{\hskip .11em plus .33em minus .07em}
 \sloppy
 \sfcode`\.=1000\relax}
\let\endthebibliography=\endlist
%
\def\lsim{\ ^<\llap{$_\sim$}\ }
\def\gsim{\ ^>\llap{$_\sim$}\ }
\def\r2{\sqrt 2}
\def\T3f{T_{3f}}
\def\sw2{\sin^2\theta_W}
\def\tw{\tan\theta_W}
\def\v#1{v_#1}
\def\tb{\tan\beta}
\def\s2b{\sin 2\beta}
\def\c2b{\cos 2\beta}
\def\s2b2{\sin^22\beta}
\def\csthR{\cos\theta_R}
\def\snthR{\sin\theta_R}
\def\csthL{\cos\theta_L}
\def\snthL{\sin\theta_L}
\def\csthf{\cos\theta_f}
\def\snthf{\sin\theta_f}
\def\sf{\tilde f}
\def\sfp{\tilde {f'}}
\def\sq{\tilde q}
\def\su{\tilde u}
\def\sd{\tilde d}
\def\sl{\tilde\ell}
\def\se{\tilde e}
\def\sn{\tilde\nu}
\def\wi{\omega_i}
\def\xj{\chi_j}
\def\g{\tilde g}
\def\mgr{m_{3/2}}
\def\m#1{\tilde {m_#1}}
\def\mH{m_H}
\def\mc{m_{\omega}}
\def\mwi{m_{\omega i}}
\def\mw#1{m_{\omega #1}}
\def\mx{m_{\chi}}
\def\mxj{m_{\chi j}}
\def\mx#1{m_{\chi #1}}
\def\mg{m_{\g}}
\def\mf{m_f}
\def\mq{m_q}
\def\Mscf{M_{\sf}}
\def\MsfL{M_{\sf L}}
\def\MsfR{M_{\sf R}}
\def\Msfk{M_{\sf k}}
\def\Msf#1{M_{\sf #1}}
\def\Msfp{M_{\sfp}}
\def\Msfpk{M_{\sfp k}}
\def\Msq{M_{\sq}}
\def\Msqk{M_{\sq k}}
\def\CL{C_L}
\def\CR{C_R}
\def\Sf{S_f}
\def\Sfp{S_{f'}}
\def\Sq{S_q}
\def\Ff{F_f}
\def\Gf{G_f}
\def\Hq{H_q}
\def\kf{\kappa_f}
\def\kfp{\kappa_{f'}}
\def\lsim{\ ^<\llap{$_\sim$}\ }
\def\gsim{\ ^>\llap{$_\sim$}\ }
\def\r2{\sqrt 2}
\def\rmuu{\gamma^{\mu}}
\def\rmud{\gamma_{\mu}}
\def\PL{{1-\gamma_5\over 2}}
\def\PR{{1+\gamma_5\over 2}}
\def\sinW2{\sin^2\theta_W}
\def\AEM{\alpha_{EM}}
\def\v#1{v_#1}
\def\tb{\tan\beta}
\def\epem{$e^+e^-$}
\def\KK{$K^0$-$\bar{K^0}$}
\def\wi{\omega_i}
\def\xj{\chi_j}
\def\Wmu{W_\mu}
\def\Wnu{W_\nu}
\def\m#1{{\tilde m}_#1}
\def\mH{m_H}
\def\mw#1{{\tilde m}_{\omega #1}}
\def\mx#1{{\tilde m}_{\chi #1}}
\def\mwi{{\tilde m}_{\omega i}}
\def\mxj{{\tilde m}_{\chi j}}
\def\rwi{r_{\omega i}}
\def\rxj{r_{\chi j}}
\def\rfp{r_f'}
\def\Kij{K_{ij}}
\def\Fq{F(q^2)}
\def\lsim{\ ^<\llap{$_\sim$}\ }
\def\gsim{\ ^>\llap{$_\sim$}\ }
\def\r2{\sqrt 2}
\def\rmuu{\gamma^{\mu}}
\def\rmud{\gamma_{\mu}}
\def\PL{{1-\gamma_5\over 2}}
\def\PR{{1+\gamma_5\over 2}}
\def\sinW2{\sin^2\theta_W}
\def\AEM{\alpha_{EM}}
\def\v#1{v_#1}
\def\tb{\tan\beta}
\def\epem{$e^+e^-$}
\def\KK{$K^0$-$\bar{K^0}$}
\def\wi{\omega_i}
\def\xj{\chi_j}
\def\Wmu{W_\mu}
\def\Wnu{W_\nu}
\def\m#1{{\tilde m}_#1}
\def\mH{m_H}
\def\mw#1{{\tilde m}_{\omega #1}}
\def\mx#1{{\tilde m}_{\chi #1}}
\def\mwi{{\tilde m}_{\omega i}}
\def\mxj{{\tilde m}_{\chi j}}
\def\rwi{r_{\omega i}}
\def\rxj{r_{\chi j}}
\def\rfp{r_f'}
\def\Kij{K_{ij}}
\def\Fq{F(q^2)}
\begin{titlepage}

\begin{flushright}
ICRR-Report-392-97-15  \\
OCHA-PP-98  \\
\end{flushright}

\  \
\renewcommand{\thefootnote}
{\fnsymbol{footnote}}
\begin{center}
{\large {\bf Electric Dipole Moments of Neutron and Electron }}  \\
{\large {\bf in Supersymmetric Model 
\footnote{To appear in the Proceedings of the KEK
meetings on '$CP$ violation and its origin' ($1993-1997$).
}
}}
\vskip 1.0 true cm
\hspace*{8.0cm} To the memory of Yoshiki Kizukuri 
\vspace{2cm}
\renewcommand{\thefootnote}
{\fnsymbol{footnote}}
Mayumi Aoki
\footnote{Graduate School of Humanities and Sciences.}
\footnote{Research Fellow of the Japan Society 
for the Promotion of Science.}, 
Tomoko Kadoyoshi${\ }^*$, Akio Sugamoto
\\
\vskip 0.5 true cm 
{\it Department of Physics, Ochanomizu University}  \\
{\it Otsuka 2-1-1, Bunkyo-ku, Tokyo 112, Japan}  \\
\vspace{1cm}
Noriyuki Oshimo
\\
\vskip 0.5 true cm
{\it Institute for Cosmic Ray Research, University of Tokyo} \\
{\it Midori-cho 3-2-1, Tanashi, Tokyo 188, Japan}  \\
\end{center}

\vskip 2.0 true cm

\centerline{\bf Abstract}
\medskip
     The electric dipole moments (EDMs) of the neutron and the electron
are reviewed within the framework of the supersymmetric standard model
(SSM) based on grand unified theories coupled to $N$=1 supergravity.
Taking into account one-loop and two-loop contributions to the EDMs, 
we explore SSM parameter space consistent with experiments 
and discuss predicted values for the EDMs.
Implications of baryon asymmetry of our universe for the EDMs are also 
discussed.  

\vskip 0.5 true cm
\noindent 

\end{titlepage}

\newpage 

\def\lsim{\ ^<\llap{$_\sim$}\ }
\def\gsim{\ ^>\llap{$_\sim$}\ }
\def\r2{\sqrt 2}
\def\rmuu{\gamma^{\mu}}
\def\rmud{\gamma_{\mu}}
\def\PL{{1-\gamma_5\over 2}}
\def\PR{{1+\gamma_5\over 2}}
\def\T3f{T_{3f}}
\def\sinW2{\sin^2\theta_W}
\def\AEM{\alpha_{EM}}
\def\v#1{v_#1}
\def\tb{\tan\beta}
\def\c2b{\cos 2\beta}
\def\epem{$e^+e^-$}
\def\KK{$K^0$-$\bar{K^0}$}
\def\wi{\omega_i}
\def\xj{\chi_j}
\def\sf{\tilde f} 
\def\Wmu{W_\mu}
\def\Wnu{W_\nu}
\def\mgr{m_{3/2}}
\def\m#1{{\tilde m}_#1}
\def\mH{m_H}
\def\mw#1{{\tilde m}_{\omega #1}}
\def\mx#1{{\tilde m}_{\chi #1}}
\def\mwi{{\tilde m}_{\omega i}}
\def\mxj{{\tilde m}_{\chi j}}
\def\MsfL{{\tilde M}_{fL}} 
\def\MsfR{{\tilde M}_{fR}} 
\def\Msfl{{\tilde M}_{f1}} 
\def\Msfh{{\tilde M}_{f2}} 
\def\rwi{r_{\omega i}}
\def\rxj{r_{\chi j}}
\def\rfp{r_f'}
\def\Kij{K_{ij}}
\def\Fq{F(q^2)}
\def\be{\begin{equation}}
\def\ee{\end{equation}}
\def\bea{\begin{eqnarray}}
\def\eea{\end{eqnarray}}
\def\nn{\nonumber }

\section{Introduction}

     Some extension of the standard model (SM) is expected 
from various viewpoints, among which is $CP$ violation.  
The violation of $CP$ invariance is observed through the phenomena  
in the $K^0-\bar {K^0} $ system \cite{CPrev}, which can be explained by 
the Kobayashi-Maskawa (KM) mechanism of the SM. 
However, it has been suggested that baryon asymmetry 
of the universe could not be explained without 
the existence of some new source of $CP$ violation \cite{BArev}.
Therefore, various extensions of the SM have been 
proposed to account for the baryon asymmetry.  
One of the extended models having such new $CP$-violating 
sources is the supersymmetric standard model (SSM),  
which is also plausible for describing physics at the electroweak scale.

     In this report we review the electric dipole moments (EDMs) of 
the neutron  $d_n$ and the electron $d_e$
coming from the new sources of $CP$ violation
in the SSM based on grand unified theories
(GUTs) coupled to $N$=1 supergravity \cite{SUSYrev}.
These EDMs arise from one-loop diagrams in which 
the squarks $\tilde q$ or sleptons $\tilde l$ propagate 
together with the charginos $\omega_i$, 
the neutralinos $\chi_j$,
or the gluinos $\tilde g$ as shown in Fig. 1 \cite{ellis,edm,edmnew}.
The EDMs also receive contributions from two-loop diagrams containing 
an effective $CP$-violating coupling  
of the $W$ bosons and photon \cite{2wedm,wedm}.
The relevant diagram is shown in Fig. 2.  
Since in the SM the EDMs are predicted to be much smaller than 
their present experimental upper bounds,  
$|d_n|<10^{-25}e$cm $\cite{nEDM}$ and $|d_e|<10^{-26}e$cm $\cite{eEDM}$, 
they can give us an important
clue to new sources of $CP$ violation.  
The new sources of $CP$ violation in the SSM could account for the 
baryon asymmetry of the universe \cite{nelson,ba1,ba2}.  
We also discuss its implications for the EDMs. 

     The SSM has at least two new $CP$-violating phases in addition to 
the KM phase $\cite{ellis}$.
One of these phases comes from the gauge-Higgs sector and another from 
the squark-slepton sector.
If these phases are not suppressed, at the one-loop level 
the chargino contributions
to the EDMs are larger than the neutralino or the gluino contributions
in wide ranges of SSM parameter space.
Comparing the chargino contributions with the experimental constraints,  
the squark and slepton masses are predicted to be larger than 1 TeV 
\cite{edm}.
On the other hand, if the $CP$-violating phase in the gauge-Higgs sector is 
quite small
while that in the squark-slepton sector being not suppressed, the gluino 
or the neutralino contributions become dominant.
In this case the squark and slepton masses of order 100 GeV are not 
contradictory to the experimental constraints for the EDMs \cite{ba2}.

     The new phase in the gauge-Higgs sector also induces the EDM of 
the $W$-boson through the one-loop diagram mediated by charginos and
neutralinos $\cite{2wedm,wedm}$.
The $W$-boson EDM, if exists, can generally yield the neutron and the 
electron EDMs by the standard electroweak interactions 
at the one-loop level \cite{1wedm}.  
As a result the EDMs of the neutron and the electron receive  
contributions at the two-loop level. 
The EDMs by these two-loop contributions could be only smaller 
than the experimental upper bounds by one order of magnitude \cite{wedm}.  
In addition, the two-loop contributions only depend on SSM parameters  
for the gauge-Higgs sector, 
whereas the one-loop contributions depend on both the 
gauge-Higgs sector
and the squark-slepton sector. 
We can predict the values of the EDMs induced by the $W$-boson EDM 
with less uncertainly.

     In the SSM, baryon asymmetry of the universe may be generated
by the $CP$-violating phases which induce the EDMs.
Indeed, the ratio of baryon number to entropy 
consistent with its observed value, 
$\rho_B/s=(2-9)\times 10^{-11}$ \cite{PDG}, 
is obtained in reasonable ranges of
SSM parameters \cite{ba1,ba2}, if the $CP$-violating phases are
not suppressed.
For these parameter ranges the EDMs are predicted to have 
sizable values.
\begin{figure}
\begin{center}
\begin{picture}(400,100)(0,0)
\CArc(95,50)(30,0,180)
\DashCArc(95,50)(30,180,360){2}
\Line(13,50)(65,50)
\Line(125,50)(177,50)
\Text(95,90)[]{$\omega$}
\Text(95,10)[]{$\tilde f'$} 
\Text(4,50)[]{$f$}
\Text(186,50)[]{$f$}
\CArc(305,50)(30,0,180)
\DashCArc(305,50)(30,180,360){2}
\Line(232,50)(275,50)
\Line(335,50)(387,50)
\Text(305,90)[]{$\chi,$ $\tilde g$}
\Text(305,10)[]{$\tilde f$} 
\Text(214,50)[]{$f$}
\Text(396,50)[]{$f$}
\end{picture}
\caption{The Feynman diagrams for the EDM of a quark or a lepton.
      Photon lines are understood.}
\end{center}
\end{figure}
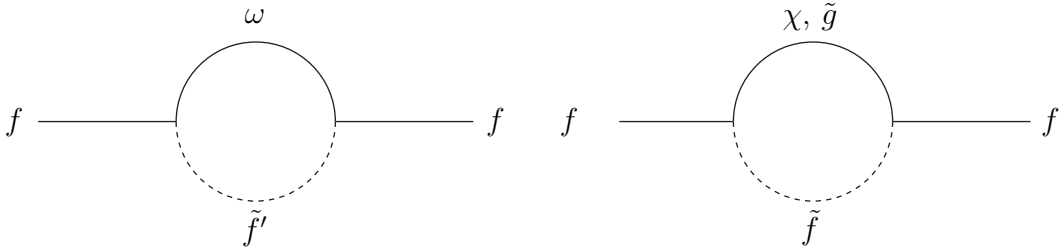
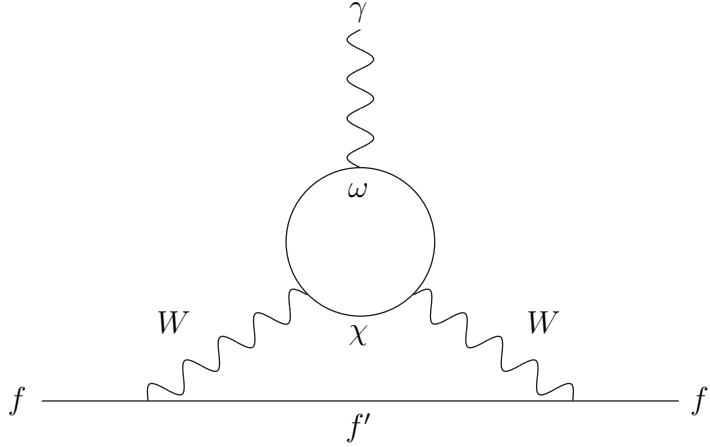
\begin{figure}
\begin{center}
\begin{picture}(400,170)(0,0)
\CArc(200,80)(28,0,360)
\Line(80,20)(320,20)
\Photon(120,20)(180,60){5}{4.5}
\Photon(280,20)(220,60){-5}{4.5}
\Photon(200,108)(200,160){5}{3.5}
\Text(200,100)[]{$\omega$}
\Text(200,10)[]{$f'$} 
\Text(71,20)[]{$f$}
\Text(329,20)[]{$f$}
\Text(200,167)[]{$\gamma$}
\Text(200,45)[]{$\chi$} 
\Text(130,50)[]{$W$}
\Text(270,50)[]{$W$}
\end{picture}
\caption{The Feynman diagram for the EDM of a quark or a lepton
which involves an effective $CP$-violating coupling for 
the $W$-bosons and photon.}
\end{center}
\end{figure}

     In Sec. 2 we summarize new origins of $CP$ violation in the SSM.
The EDMs of the quarks and the leptons can arise at the one-loop level,
which are given and numerically evaluated in Sec. 3.
The EDMs by the two-loop effects are discussed in Sec. 4.
The new sources of $CP$ violation generate the baryon asymmetry, 
whose constraints on the EDMs are shown in Sec. 5.
A summary is given in Sec. 6. 

\section{Model}

     The SSM based on GUTs coupled to $N$=1 supergravity has 
several complex
parameters in addition to the Yukawa coupling constants.
In the model with minimal particle contents, these complex parameters 
which are
possible new sources of $CP$ violation are 
the SU(3), SU(2), and U(1) gaugino masses $\tilde m_3,$ $\tilde m_2,$ and 
$\tilde m_1,$
respectively, the mass parameter $m_H$ 
in the bilinear term of Higgs superfields in superpotential,
and the dimensionless parameters $A_f$'s and $B$ in the trilinear and bilinear 
terms of scalar fields.

     The complex parameters lead to complex mass terms 
for supersymmetric particles.
When the SU(2)$\times$U(1) gauge symmetry is broken, the mass matrices 
$M^-$ and $M^0$
for the charginos and the neutralinos become 
\bea
    M^- &=& \left(\matrix{\m2 & -g\v1^*/\r2 \cr
                -g\v2^*/\r2 & \mH}        \right),      
\label{cha mass} \\
    M^0 &=& \left(\matrix{\m1 &  0  & g'\v1^*/2 & -g'\v2^*/2 \cr
                         0  & \m2 & -g\v1^*/2 &   g\v2^*/2 \cr
                       g'\v1^*/2 & -g\v1^*/2 &   0  & -\mH \cr
                      -g'\v2^*/2 &  g\v2^*/2 & -\mH &   0}
           \right),   
\label{neu mass}
\eea
where $v_1$ and $v_2$ are the vacuum expectation values
of the two Higgs doublets with U(1) hypercharges 
$-1/2$ and $1/2$, respectively.
These mass matrices are diagonalized by unitary 
matrices $C_R$, $C_L$, and $N$ as
\bea
      \CR^{\dag}M^-\CL &=& {\rm diag}(\mw1, \mw2),      \\
   N^tM^0N &=& {\rm diag}(\mx1, \mx2, \mx3, \mx4),
\eea
giving the mass eigenstates.
For the mass parameters of the gauginos we assume the relation
$(g^2/g_s^2)\tilde m_3=\tilde m_2=(3g^2/5g'^2)\tilde m_1$ suggested by
GUTs.

     For the squarks there are two species for each flavor,
the left-handed squark $\tilde q_L$ and
the right-handed squark $\tilde q_R$,
corresponding to two chiralities of the quark.  
There also exist sleptons $\tilde l_L$ and $\tilde l_R$ similarly 
corresponding to the leptons.  
The mass-squared matrix for the squarks or sleptons with 
flavor $f$ becomes
\be
  \lefteqn{\tilde M^2_f = }\hspace{8 cm}  
 \label{sq mass}   
\ee
\[
    \left(\matrix{\mf^2 + \c2b (\T3f - Q_f\sw2 )M_Z^2 + \MsfL^2 &
                                            \mf (R_f\mH + A_f^*\mgr) \cr
                   \mf (R_f^*\mH^* + A_f\mgr) &
                               \mf^2 + Q_f\c2b \sw2 M_Z^2 + \MsfR^2}
           \right), 
\]
\bea
   R_f &=& 
               {\v1 \over\v2^*}\quad(\ \T3f =   {1\over 2}\ ), 
  \quad    {\v2 \over\v1^*}\quad(\ \T3f = - {1\over 2}\ ),  \nn \\
   \tb &=& \left|{\v2 \over\v1}\right|.  \nn 
\eea
Here $m_f$ represents a mass of the fermion $f$, $Q_f$ an electric charge of
$f$, and $T_{3f}$ the third component of the weak isospin for the left-handed
component of $f$. 
The gravitino mass is denoted by $m_{3/2}$, and $M^2_{\tilde fL}$ and
$M^2_{\tilde fR}$ are mass-squared parameters for $\tilde f_L$ and $\tilde 
f_R$, respectively.
Each mass-squared matrix for the squarks and the sleptons is diagonalized by 
a unitary matrix $S_f$ as
\be
      \Sf^{\dag}\tilde M^2_f\Sf = {\rm diag}(\Msf1^2, \Msf2^2).
\ee
We have neglected generation mixings.  

     All the complex phases of the parameters in Eqs. (\ref{cha mass}), 
(\ref{neu mass}), and (\ref{sq mass}) are not physical.
By the redefinition of the fields we can take without loss of generality 
$\tilde m_i$ $(i=1-3),$ $v_1,$ and $v_2$ for real and positive.
Then the remaining parameters $m_H$ and $A_f$'s cannot be made real, 
which are origins of $CP$ violation.
We express them as 
\bea
  m_H &=& |m_H|\exp(i\theta),      \\
  A_f &=& A = |A|\exp(i\alpha).  
\eea
Since $A_f$'s are considered to have the same value of order unity
at the grand unification scale, their differences at the electroweak
scale are small and thus can be neglected.

\section{EDM at one-loop level}

     The EDM of the quark receives contributions  
at the one-loop level
from diagrams in which the charginos, neutralinos or gluinos are exchanged
together with the squarks as shown in Fig. 1.
The electron EDM is also induced by one-loop diagrams, where the charginos or 
neutralinos are exchanged together with the sleptons.

     The EDM operator changes the chirality of the quark or the lepton.
The gauginos couple the quark (lepton) to the squark (slepton)
with the same chirality via the gauge interactions, while 
the Higgsinos couple 
the quark (lepton) to the squark (slepton) with the opposite chirality via 
the Yukawa interactions.
Therefore the flip of the chirality can arise at the one-loop level 
from three origins as 
follows.

(i) One vertex of the loop diagram is a gauge interaction and the 
other is a Yukawa interaction.
The gaugino and the Higgsino are mixed.

(ii) The two vertices are both gauge interactions.
The scalar particles $\tilde q_L$ $(\tilde l_L)$ and 
$\tilde q_R$ $(\tilde l_R)$ are mixed.

(iii) The two vertices are both Yukawa interactions.
The scalar particles are mixed. 

\noindent   
The chargino-loop diagram originates in (i) and (iii), 
the gluino-loop diagram in (ii),
and the neutralino diagram in all of these.

     The relative magnitudes of the chargino, neutralino, 
and gluino contributions are crudely estimated by 
considering their origins.  
The gaugino-Higgsino mixing and the $\tilde q_L$-$\tilde q_R$ or 
$\tilde l_L$-$\tilde l_R$ mixing are respectively suppressed by roughly 
$M_W/\tilde m_\omega$ and $m_f/M_{\tilde f}$.
The products for the coupling constants and suppression factor
in the origins (i), (ii), and (iii) become  
$\alpha m_f/\tilde m_\omega$, $\alpha m_f/M_{\tilde f}$, and
$\alpha m_f^3/M_W^2 M_{\tilde f}$, respectively, 
$\alpha$ being an appropriate fine structure constant.
Therefore the EDM from the chargino contribution is proportional
to $\alpha_2 m_f/\tilde m_\omega$,
while that from the gluino contribution is proportional to  
$\alpha_s m_f/M_{\tilde f}$.
If the squark and chargino masses satisfy the condition
$M_{\tilde f} > (\alpha_s/\alpha_2)\tilde m_\omega$,
the chargino contribution becomes larger than the gluino contribution.
The neutralino contribution is the smallest 
among all the contributions and may be neglected, 
since the coupling strength of the neutralino is smaller than those of
the gluino and the chargino. 

\begin{figure}
\vspace{9cm}
\includegraphics{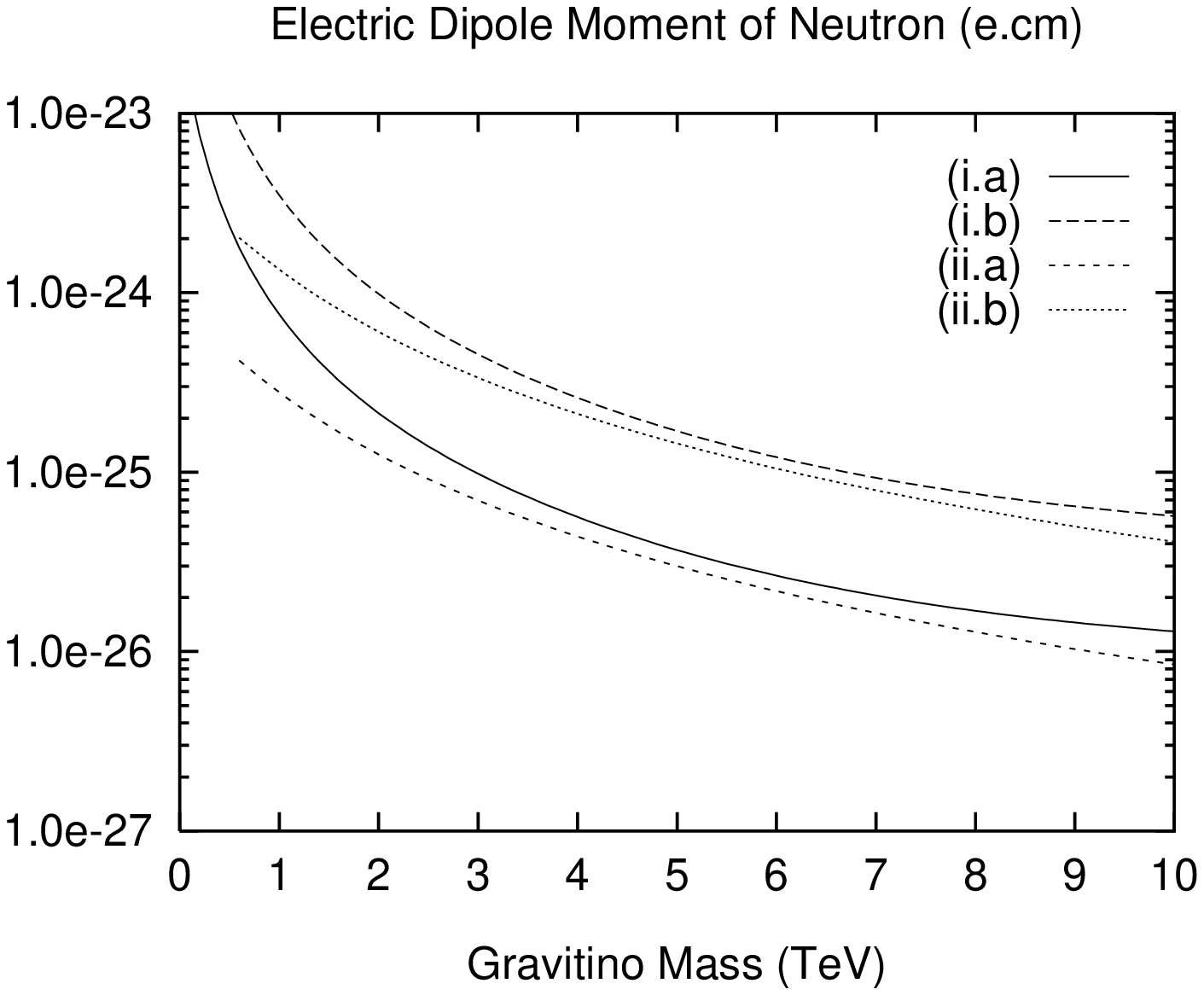}
\caption{The absolute value of the neutron EDM from the chargino contribution
for the parameter values given in Table 1.}
\label{fig3}
\end{figure}         
\begin{table}
\begin{center}
\begin{tabular}{lcccc}
\hline
      & (i.a) & (i.b) & (ii.a) &(ii.b) \\
\hline
 $\m2=|\mH|$ (GeV) & 200 & 200 & 1000 & 1000 \\
$\tb$              & 2 & 10  & 2 & 10  \\
\hline
\end{tabular}
\end{center}
\caption{The values of $\m2=|m_H|$ and $\tb$ for curves 
            (i.a)--(ii.b) in Fig. 3.} 
\label{tab1}
\end{table}

     The EDM of a quark or a lepton from the chargino contribution 
is given by 
\bea
     d_f^C/e &=& {\alpha_{EM}\over 4\pi\sw2}R_f\sin\theta
     {\m2 |\mH|\over (\mw2^2 - \mw1^2)}{\mf\over\Msfp^2} \nn \\
        & & \times\sum_{i=1}^2 (-1)^i
        [Q_{\sfp}I\left({\mwi^2\over\Msfp^2}\right) +
        (Q_f - Q_{\sfp})J\left({\mwi^2\over\Msfp^2}\right)], 
\label{chargino con} \\ 
I(r) &=& {1\over 2(1 - r)^2}\left(1 + r + {2r\over 1 - r}\ln r\right), \nn\\
J(r) &=& {1\over 2(1 - r)^2}\left(3 - r + {2\over 1 - r}\ln r\right). \nn 
\eea
Here we have neglected the tiny contribution from the origin (iii) and
approximated the masses of two mass eigenstates of the
squarks or the sleptons to have the same value 
$\Mscf$ $(\equiv \Msf1\simeq\Msf2)$. 
For a crude estimate of the chargino contribution,       
we note that the numerical value of a factor 
in Eq. (\ref{chargino con}) is written by  
\be
{\alpha_{EM}\over 4\pi\sw2}{\mf\over\Msfp^2} =
5.0\times 10^{-25}\left({1~{\rm TeV}\over\Msfp}\right)^2
\left({\mf\over 10~{\rm MeV}}\right)\ {\rm cm},
\ee
and $I(r)$ and $J(r)$ vary as $5\times 10^{-1}-5\times 10^{-3}$ 
and $-3-(-5)\times 10^{-3}$, respectively, 
for $10^{-2} < r < 10^2$.  
If the $CP$-violating phase $\theta$ is not suppressed,
the squark or slepton masses or the
chargino masses have to be larger than 1 TeV for satisfying  
the experimental constraints on the EDMs of the neutron and the electron.   

     The EDM of a quark from the gluino contribution is given by
\bea
     d_q^G/e &=& {2\alpha_s\over 3\pi}\left(\sin\alpha |A|
          - R_q\sin\theta{|\mH |\over\mgr}\right){\mgr\over\Msq}
  {\mq\over\Msq^2}Q_{\sq}{\mg\over\Msq}K\left({\mg^2\over\Msq^2}\right), \\
K(r) &=& {-1\over 2(1 - r)^3}[1 + 5r +
                             {2r(2 + r)\over 1 - r}\ln r],  \nn 
\eea
where $\Msq$ denotes the average of $M_{\tilde q1}$ and $M_{\tilde q2}$.  
Although the factor $2\alpha_s/3\pi$ is about ten times
larger than the factor $\alpha_{EM}/4\pi\sw2$,
the function $\sqrt r K(r)$ takes a value
about ten times smaller than $I(r)$ or $J(r)$, 
$-1\times 10^{-1}<{\sqrt r}K(r)<-2\times 10^{-3}$, for
$10^{-2} < r < 10^2$. 
Therefore, the constraints from the gluino contribution are  
similar to those from the chargino contribution.  

\begin{figure}
\vspace{9cm}
\includegraphics{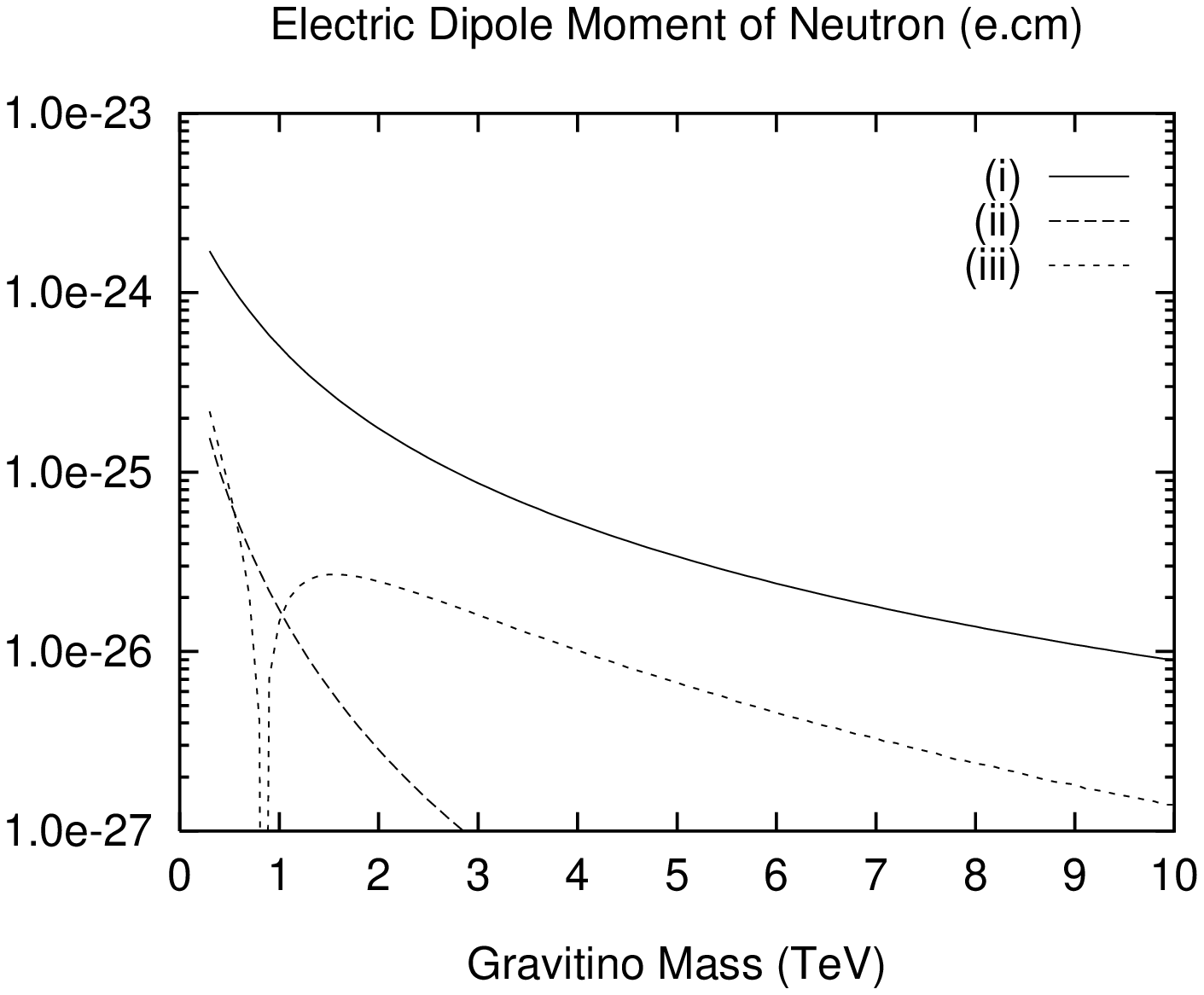}
\caption{The three contributions
to the neutron EDM for $\tb = 2$ and $\m2 = |\mH| = 500$ GeV:
(i)chargino, (ii)neutralino, (iii)gluino.   
The absolute values of the EDM are shown.}
\label{fig4}
\end{figure}         

     We evaluate the supersymmetric contributions to the EDM of the neutron.
Assuming the nonrelativistic quark model, the EDMs of the $u$ and $d$
quarks are converted into the EDM of the neutron: $d_n=(4d_d-d_u)/3$.
The mass parameters $M_{\sf L}$ and $M_{\sf R}$ appearing 
in Eq. (\ref{sq mass})
could be related to the gravitino mass and the gaugino masses.  
In the ordinary scheme for the mass generation,
the gaugino masses are smaller than or around the gravitino mass,
so that a scale characteristic of $M_{\sf L}$ and $M_{\sf R}$ is
given by $\mgr$.  For simplicity,
we take $M_{\su L}=M_{\su R}=M_{\sd L}=M_{\sd R}=\mgr$.
Then the masses of the squarks
can be estimated approximately by $\mgr$.  The magnitude of the
Higgsino mass parameter $|m_H|$ should be at most of order $\mgr$ for
correctly breaking the SU(2)$\times$U(1) symmetry.
As a typical example for a natural magnitude of the 
$CP$-violating phases,
we simply take $\theta =\alpha =\pi/4$.
The absolute value of the dimensionless parameter $A$ is fixed as $|A|=1$. 

\begin{figure}
\vspace{9cm}
\includegraphics{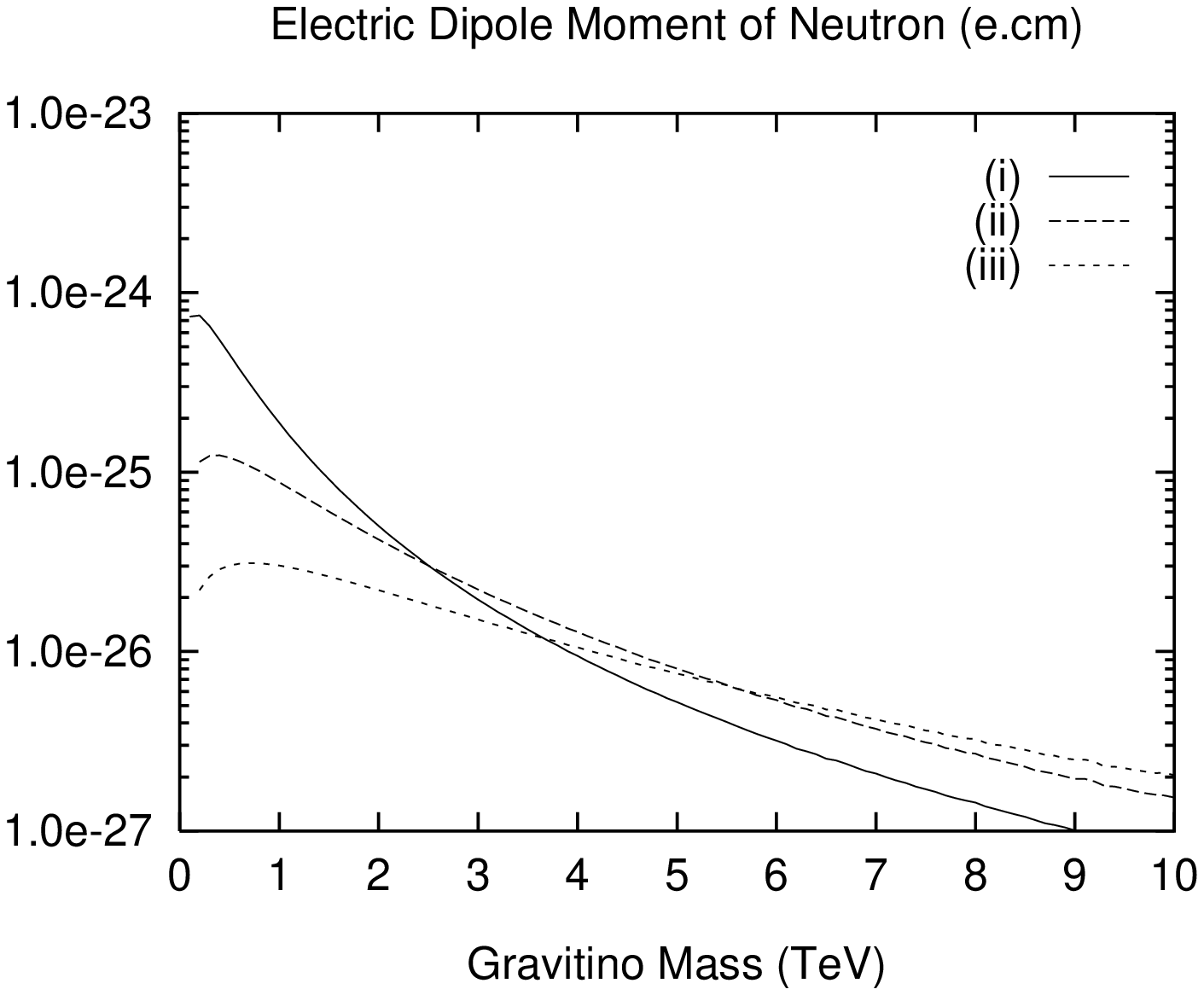}
\caption{The neutron EDM as a function of $\mgr$ for $\alpha=\pi/4$
and $\theta=0$.
Three curves correspond to three values for $\m2$ : (i) 200 GeV, 
(ii) 500 GeV, (iii) 1 TeV.
The other parameters are taken for $\tb$=2 and $|m_H|=200$ GeV. }
\label{fig5}
\end{figure}         

     In Fig. 3 the absolute value of the chargino contribution 
to the EDM of the neutron is plotted 
as a function of $m_{3/2}$.
The values of the mass parameters and $\tb$ are given in Table 1.
In the ranges of $\mgr$ where curves are not drawn, the lightest squark is
lighter than the lightest neutralino, which is disfavored by cosmology.
In the presented parameter regions the gluino and the neutralino
contributions are smaller than the chargino contribution, 
so that the EDM of the neutron is given by $d_n\simeq d_n^C$.  
Not to be conflict
with the experimental upper bound of $|d_n|<10^{-25}e$cm, we must have
$\mgr\gsim 3 $ TeV for $\tb=2$ and $\mgr\gsim 7 $ TeV for $\tb=10$.
The clear dependence on $\tb$ arises from the dominance of the 
$d$-quark EDM over the $u$-quark EDM. Since the former 
is proportional to $\tb$, the neutron EDM increases as 
$\tb$ becomes large.  

     The difference between the three contributions of   
the chargino, the neutralino, and the gluino can be seen in
Fig. 4, where $|d_n^C|,$ $|d_n^N|$, and $|d_n^G|$ are shown as
functions of $\mgr$ for $\tb =2$ and $\m2 =|\mH |=500$ GeV.
The sign of the neutron EDM from the gluino contribution changes 
at $\mgr \approx 800-900$ GeV
owing to the interference of the two $CP$-violating phases 
$\theta$ and $\alpha$.
This figure clearly shows that the charginos really give
the largest contribution to $d_n$. Since $d_n^C$ is dominant,
the EDM of the neutron is roughly proportional 
to $\sin\theta$ as seen from Eq. (\ref{chargino con}),
and does not depend much on $\alpha$.

     The electron EDM induced through one-loop diagrams has 
a value smaller than the neutron EDM by one order of magnitude.
This difference comes simply from the difference between the 
masses of the electron and the $d$ quark.  The electron EDM 
satisfies the constraint $|d_e|<10^{-26}e$cm in the ranges 
$\mgr\gsim 1$ TeV and $\mgr\gsim 4$ TeV for $\tb=2$ 
and $\tb=10$, respectively.  
Similarly to the neutron EDM, the electron EDM increases as 
$\tb$ becomes large.  

     The EDMs of the neutron and the electron decrease as 
the phases $\theta$ and $\alpha$ become small.  
If $\theta$ is much smaller than $\alpha$, 
the EDMs of the neutron and the electron receive dominant 
contributions from the gluino-loop and the neutralino-loop diagrams,
respectively.
Then the constraints of the EDMs on the SSM parameters become
relaxed.  
In Fig. 5 the neutron EDM is shown as a function of the gravitino mass
for $\alpha = \pi/4$ and $\theta=0$, taking three values for $\m2$ :
(i) 200 GeV, (ii) 500 GeV, (iii) 1 TeV.
The other parameters are fixed as $\tb$=2 and $|m_H|=200$ GeV.
In the ranges of the gravitino mass where curves are not drawn, 
the lightest squark is lighter than either 45 GeV or the lightest
neutralino.  For 500 GeV$\lsim \tilde m_2$ 
the magnitude of the neutron EDM is below the experimental upper bound
even if $\mgr$ is of order 100 GeV.   

\section{EDM at two-loop level}

     The neutron and the electron EDMs 
are generated at the two-loop level 
through the $W$-boson EDM which is induced by the  
one-loop diagram mediated by the charginos and neutralinos 
as shown in Fig. 2.
At the two-loop level there also exist diagrams 
which involve squarks or 
sleptons and make contributions to the quark or lepton EDM.  
However,
as long as the squarks and sleptons are much heavier than 
the charginos and neutralinos, these diagrams can be safely 
neglected.  

     The EDM of a quark or a lepton from the two-loop diagram is given by
\bea
 d_f&=&\mp e\left(\frac{\AEM}{4\pi\sinW2}\right)^2
	 \sum_{i=1}^2\sum_{j=1}^4
{\rm Im}(G_{Lji}^*G_{Rji})\frac{\mwi\mxj}{M_W^2}\frac{m_f}{M_W^2} 
  \nonumber \\
& & \frac{1}{2(1-\rfp)^2}\int_{0}^{1}\frac{ds}{s}
  [\{\frac{3-\rfp}{1-\rfp}\frac{1}{\Kij-\rfp}
                +\frac{1}{(\Kij-\rfp)^2}\}\rfp^2\ln\frac{\rfp}{\Kij} 
  \nonumber \\
& & +\{\frac{1-3\rfp}{1-\rfp}\frac{1}{\Kij-1}
        +\frac{\rfp}{(\Kij-1)^2}\}\ln\frac{1}{\Kij} \nonumber \\
& & +\left(\frac{1}{\Kij-\rfp}+\frac{1}{\Kij-1}\right)\rfp], 
\label{2loop con} \\ 
           \Kij &=& \frac{\rwi}{s}+\frac{\rxj}{1-s}, \nn \\  
 \rfp &=& \frac{m_{f'}^2}{M_W^2}, 
 \quad \rwi=\frac{\tilde m_{\omega i}^2}{M_W^2}, 
\quad \rxj=\frac{\tilde m_{\chi j}^2}{M_W^2}, \nn 
\eea
where the coefficients $G_L$ and $G_R$ are defined as 
\bea
  G_{Lji} &=& \sqrt{2}N_{2j}^*C_{L1i}+N_{3j}^*C_{L2i},  \nn \\ 
  G_{Rji} &=& \sqrt{2}N_{2j}C_{R1i}-N_{4j}C_{R2i}. 
\eea
The negative and positive signs of Eq. (\ref{2loop con}) 
are, respectively, for the fermions with the 
weak isospins 1/2 and $-1/2$.  
This two-loop contribution only depends on the parameters 
$\tb$, $\m2$, and $\mH$ contained in the gauge-Higgs sector.  

\begin{figure}
\vspace{9cm}
\includegraphics{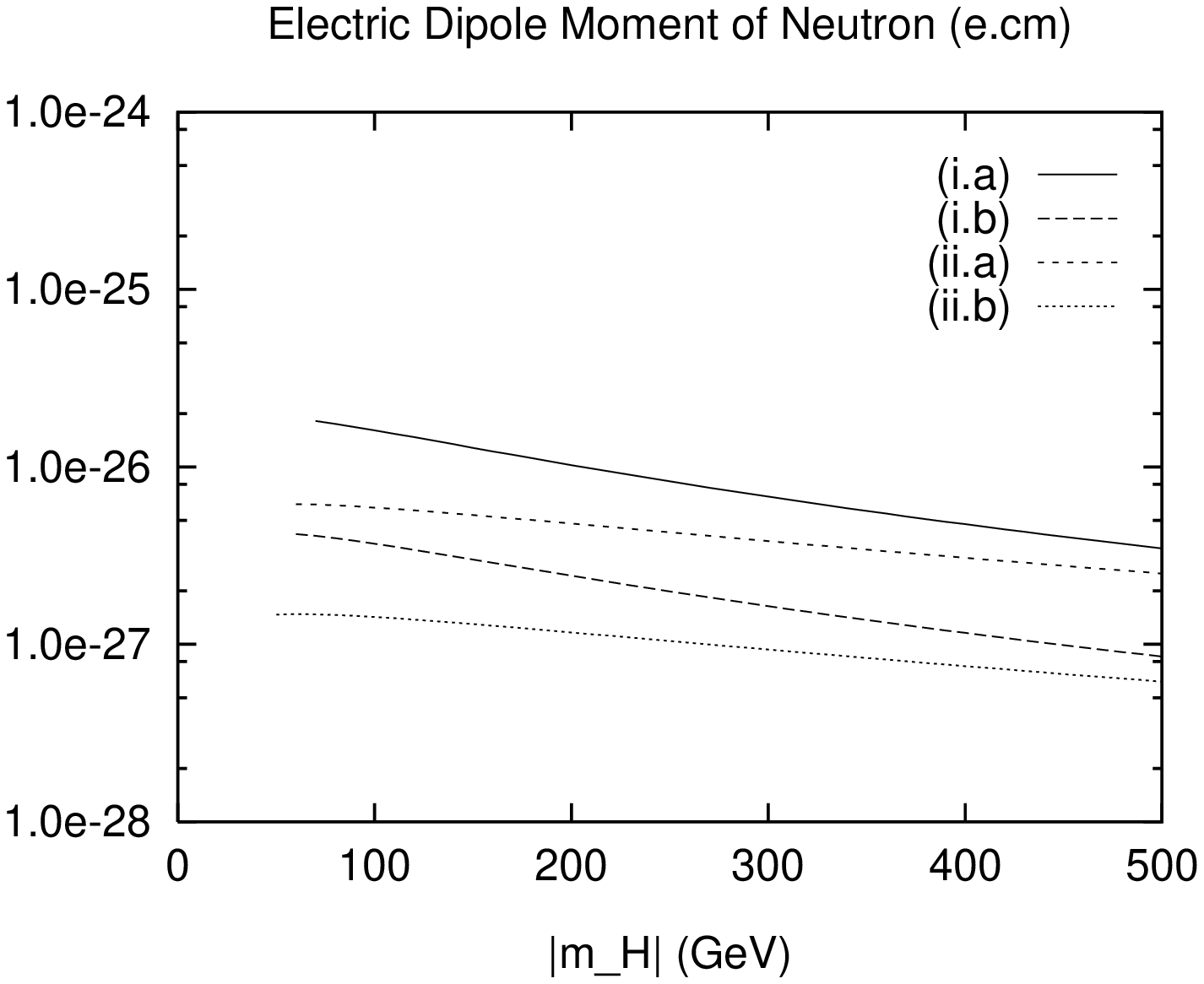}
\caption{The absolute value of the neutron EDM  
induced by the $W$-boson EDM for
$\theta=\pi/4$. The values of $\m2$ and $\tb$ for curves (i.a)--(ii.b)
are given in Table 2.} 
\label{fig6}
\end{figure}         
\begin{table}
\begin{center}
\begin{tabular}{lcccc}
\hline
    & (i.a) & (i.b) & (ii.a) &(ii.b) \\
\hline
 $\m2$ (GeV) & 200 & 200 & 500 & 500 \\
 $\tb$  & 2  & 10  & 2 & 10  \\
\hline
\end{tabular} 
\end{center}
\caption{The values of $\m2$ and $\tb$ for curves 
            (i.a)--(ii.b) in Fig. 6.} 
\label{tab2}
\end{table}

     In Fig. 6 the absolute value of  
the neutron EDM induced by the $W$-boson EDM is shown as a function 
of $|\mH|$. 
For $\m2$ and $\tb$ we have taken four sets of values given in 
Table 2.  The $CP$-violating phase is fixed as $\theta=\pi/4$.
In the ranges of $|m_H|$ where curves are not drawn, the lighter chargino
has a mass smaller than 45 GeV which has been ruled out by LEP
experiments $\cite{PDG}$.
For $\m2,$ $|\mH|\sim 100$ GeV and $\tb\sim 1$, 
the magnitude of the neutron EDM is around 
$10^{-26} e$cm, which is smaller than the present 
experimental upper bound by only one order of magnitude.  
The EDM of the neutron decreases as $\m2$ or $|\mH|$ increases.  
Since the squarks have masses larger than 1 TeV for $\theta\sim 1$, 
the contributions to the neutron EDM from two-loop diagrams 
mediated by the squarks are negligible.
The EDM of the electron induced by the $W$-boson EDM has a value 
smaller than the neutron EDM by one order of magnitude, similarly
depending on the SSM parameters. 

     The neutron and electron EDMs at the two-loop level 
do not vary with the squark and slepton masses if these 
masses are enough heavy.  
On the other hand, the one-loop contributions to the EDMs decrease, 
as the squark and slepton masses become large.   
If the squarks and sleptons have masses around 10 TeV,
the two-loop contributions become comparable with the one-loop contributions.  
Furthermore, it turns out that these one-loop and two-loop 
contributions to the EDM of the neutron or the electron have the same sign.  
Therefore, the neutron and electron EDMs arising from the 
two-loop diagrams give theoretical lower bounds for 
given parameter values of the gauge-Higgs sector.   

\section{Constraints from baryon asymmetry}

     The $CP$-violating phase $\theta$  
could induce baryon asymmetry of the universe
through the charge transport mechanism \cite{nelson2} mediated by 
the charginos.
Assuming that the electroweak phase transition of the universe is first order, 
bubbles of the broken phase nucleate in the symmetric phase.  
The charginos incident on the bubble wall from
the symmetric phase or the broken phase are 
reflected or transmitted to the symmetric phase.   
In these processes $CP$ violation makes differences in reflection
or transmission probability between a particle state and its 
$CP$-conjugate state, leading to a net density of hypercharge.  
Equilibrium conditions in the symmetric 
phase are then shifted to favor a non-vanishing value for baryon 
asymmetry, which is realized through electroweak anomaly.
The $CP$-violating phase $\alpha$ also enables the  
$t$ squarks to assume the role of the mediator for the 
charge transport mechanism.  

     An enough amount of baryon asymmetry can be induced by the chargino
transport, if the phase $\theta$ is not suppressed and the chargino
masses are of order 100 GeV.
In this case the squark and slepton masses are larger than 1 TeV.   
The neutron EDM is then predicted to be $10^{-25}-10^{-26}e$cm for 
1 TeV $\lsim M_{\tilde q} \lsim$ 10 TeV as shown in Fig. 3.
If $\theta$ is much smaller than unity while $\alpha$ being not 
suppressed, the $t$ squarks with their 
masses of order 100 GeV can generate an enough amount of the asymmetry.
In this case the masses of the other squarks and sleptons 
are also of order 100 GeV.
The EDM of the neutron is predicted to be  $10^{-25}-10^{-26} e$cm for 
500 GeV $\lsim \m2 \lsim$ 1 TeV as shown in Fig. 5.  
If the baryon asymmetry originates in the new sources of $CP$ 
violation in the SSM, the neutron EDM has a magnitude 
which can be explored in the near future.  
The electron EDM is also predicted not to be much smaller than its 
experimental upper bound.
 
\section{Summary}

     We have discussed the EDMs of the neutron and 
the electron in the SSM,
under the assumption of GUTs and $N$=1 supergravity. 
The SSM has two $CP$-violating phases $\theta$ and $\alpha$ 
intrinsic in the model.  
These EDMs receive contributions at the one-loop level from 
the diagrams in which the charginos, neutralinos or gluinos are
exchanged together with the squarks or sleptons.
Among these contributions the chargino contribution dominates over  
the gluino and neutralino contributions  
in wide ranges of the parameter space.   
If the $CP$-violating phase $\theta$ is of order unity, 
the experimental constraints on the neutron and the electron EDMs 
give the prediction that the squarks and sleptons are 
heavier than 1 TeV.

     The EDMs also receive contributions at the two-loop level induced by 
the $W$-boson EDM. 
If the squarks and sleptons are much 
heavier than the charginos and neutralinos, 
other two-loop diagrams 
with the squarks or sleptons are neglected,  
which may be indeed the case for $\theta$  
of order unity and thus  
the squark and slepton masses larger than 1 TeV.  
Since the two-loop contributions by the $W$ boson EDM do not 
depend on the squark or slepton masses, 
the resultant values of the EDMs are less ambiguous than 
the one-loop contributions.  
The neutron and the electron EDMs induced at the 
two-loop level are around $10^{-26}e$cm and $10^{-27}e$cm, 
respectively, 
for the chargino and the neutralino masses of order 100 GeV.

     We have also discussed the relation between the EDMs and baryon 
asymmetry of the universe.
If $\theta$ or $\alpha$ are not much suppressed, 
the charginos or the $t$ squarks with their masses of order 100 GeV
can mediate the charge transport mechanism to generate 
the asymmetry consistent with its observed value.
Then it is likely that the neutron and electron EDMs 
have magnitudes of
$10^{-25}-10^{-26}e$cm and $10^{-26}-10^{-27}e$cm, respectively.  
These numerical outcomes are not so small compared to the experimental 
upper bounds at present, 
and thus may be accessible in near future experiments.  

\section*{Acknowledgments} 
     N.O. is grateful to K. Hagiwara for constant encouragement extended 
to him.     
The work of M.A. is supported in part by the Grant-in-Aid 
for Scientific Research from the Ministry of Education, Science
and Culture of Japan.
This work is supported in part by the Grant-in-Aid for Scientific 
Research (No. 08640357) and by the Grant-in-Aid for joint Scientific 
Research (No. 08044089) from the Ministry of Education, Science and
Culture, Japan.

\newpage

\end{document}